\documentclass{kapproc} 
\newcommand{\be}{\begin{equation}}
\newcommand{\ee}{\end{equation}}
\newcommand{\bea}{\begin{eqnarray}}
\newcommand{\eea}{\end{eqnarray}}

\def\Journal#1#2#3#4{{#1} {\bf #2}, #3 (#4)}

\def\NPB{{\em Nucl.~Phys.} B}
\def\NPA{{\em Nucl.~Phys.} A}
\def\PLB{{\em Phys.~Lett.}  B}
\def\PRL{\em Phys.~Rev.~Lett.~}
\def\PRD{{\em Phys.~Rev.} D}
\def\PRC{{\em Phys.~Rev.} C}

\def\PRep{\em Phys.~Rep.~}

\usepackage{procps}

\usepackage[dvips]{graphicx}

\upperandlowercase

\nochapequationnumber 

\let\footnote\savefootnote

\let\footnotetext\savefootnotetext

\normallatexbib 

\begin{document}



\articletitle [Aspects of Non-Equilbrium Quantum Field Theory
in Heavy Ion Collisions]{Aspects of Non-Equilbrium Quantum Field Theory
in relativistic Heavy Ion Collisions}










--------------



\author{C.~Greiner\altaffilmark{1}, S.~Juchem\altaffilmark{2}
and Z.~Xu\altaffilmark{1,2}}

\altaffiltext{1}{Institut f\"ur Theoretische Physik der Johann Wolfgang Goethe
Universit\"at,
 D-60054, Frankfurt am Main, Germany}
\altaffiltext{2}{Institut f\"ur Theoretische Physik der Justus Liebig
Universit\"at,
 D-35392, Giessen, Germany}




\begin{abstract}
In this lecture we review recent progress in various aspects
of non-equilibrium QFT with respect to relativistic heavy ion collisions.
As a first and rather general study we summarize our (numerical)
investigations for a dissipative quantum time evolution of $\phi^4$-field 
theory
for a spatially homogeneous system in 2+1 space-time 
dimensions on the basis of the Kadanoff-Baym equations.
The initial conditions can be chosen arbitrarily and far from equilibrium.
The calculations demonstrate how thermalization and chemical 
equilibration is achieved ab initio from the underlying QFT.
As a possible candidate for a non-equilibrium phase transition phenomena
we then adress the stochastic formation of disoriented chiral condensates.
Finally we elaborate on a new 3+1 dimensional Monte Carlo parton cascade
solving kinetic  Boltzmann processes including inelastic multiplication
processes ($gg\leftrightarrow ggg$)
in an unified manner, where the back reaction channel is treated 
for the first time fully consistently.

\end{abstract}


\section{Introduction, overview and summary}

The prime intention for present ultrarelativistic heavy ion collisions
at CERN and at Brookhaven lies in the possible experimental identification of
the quark gluon plasma (QGP).
The evolving system in the reaction is at least initially
far from any (quasi-)equilibrium configuration.
Not only for this reason,
nonequilibrium quantum field theory has
become a major topic of research for describing  transport
processes in various areas of physics like in cosmological particle physics
as well as condensed matter physics. The multidisciplinary aspect
arises due to a common interest to understand the various
real-time relaxation phenomena of quantum dissipative systems.

A seminal work in solving numerically for the first time
the so called Kadanoff-Baym equations of evoltion equations for 
out-of-equilibrium
one-particle Green's function
has been carried out by Danielewicz \cite{dan84b}, who
investigated a spatially homogeneous system with a deformed Fermi sphere in
momentum space for the initial distribution of occupied momentum
states in order to model the initial condition of a heavy-ion
collision in the nonrelativistic domain. In comparison to a
standard on-shell semi-classical Boltzmann equation the solutions
showed quantitative differences like
a larger collective relaxation time for 
achieving complete kinetic equilibration.
Similar quantum modifications in the
equilibration and momentum relaxation have been found
for a relativistic situation in \cite{CGreiner}, where also a faster
equilibration can occur at higher bombarding energies.

In section 2 we summarize our findings
of a recent and very detailed study concerning the quantum time
evolution of $\phi^4$-field theory
for a spatially homogeneous system in 2+1 space-time 
dimensions by including the 
mean-field (tadpole) and collisional (sunset) self-energies \cite{Ju03}.
As an important and new step in the field, we point out
the dynamics of the spectral (`off-shell')
distributions of the excited quantum modes and the different phases 
in the approach to equilibrium described by Kubo-Martin-Schwinger (KMS) 
relations for full thermal equilibrium states, the latter being
the only translation
invariant solution and representing the stationary fixed points of the
dissipative coupled equation of motions.
Concentrating further on the importance of off-shell effects,
far off-shell $1\leftrightarrow 3$ decay processes are  responsible 
for chemical
equilibration. The physical process corresponds to a consistent 
treatment of Bremsstrahlung type processes in the medium
by the interplay of production, propagation and 
subsequent decay of highly virtual mode
excitations.

\begin{figure}[htb]
\begin{center}
\includegraphics[width=11cm]{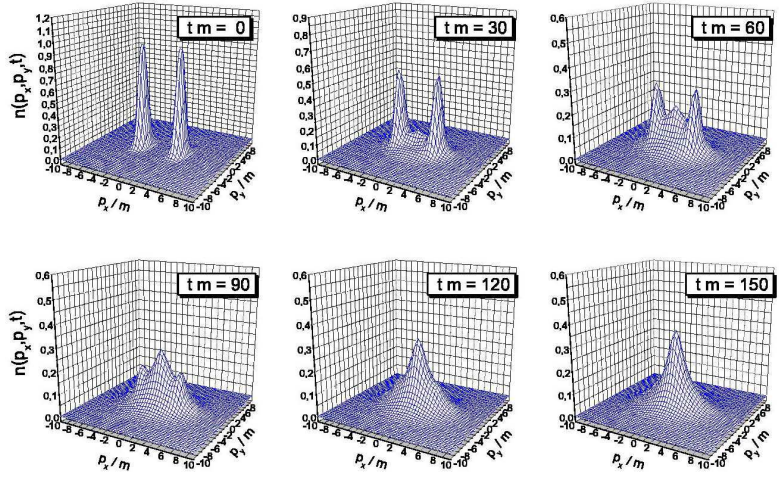}
\end{center}
\caption{\label{fig:3d1}
Characteristic evolution of the occupation number 
in momentum space within the full
Kadanoff-Baym dynamics
starting from an initially non-isotropic shape.
The occupation number correspondes directly to the 
equal-time Green function via the relation \protect\ref{ew}
and is displayed for various times 
$\bar{t} \cdot m =$ 0, 30, 60, 90, 120, 150. }
\end{figure}

As an exotic candidate for a direct signature
stemming from the QGP phase transition
(and which can be adressed in RHIC experiments at Brookhaven)
we summarize in section 3 ideas of stochastic formation
of so called disoriented chiral condensates (DCC).
The idea of DCC \cite{DCC} first appeared in a work of Anselm
but it was made widely known due to Bjorken, and Rajagopal and Wilczek.
The spontaneous growth and subsequent
decay of these configurations emerging after a rapid
chiral phase transition from the QGP to the hadronic world
would give rise to large collective fluctuations
in the number of produced low momentum neutral pions compared to charged pions.
Our work concentrates on the important question on
the likelihood of an instability leading potentially to a large DCC yield
of low momentum pions \cite{Xu00}.
It reveals that an experimentally feasible DCC, if it does exist
in nature, has to be a rare event with some finite probability
following a nontrivial and nonpoissonian distribution on an event by event
basis. DCCs could then (only) be revealed experimentally by inspecting
higher order factorial cumulants $\theta _m$ ($m\ge 3$)
in the sampled distribution of low momentum pions.

In section 4 we finally
detail on a new parton cascade scheme
treating elastic and inelastic multiplication collisions
in an unified manner \cite{Xu04}.
For the first time the back reation channel
($ggg\rightarrow gg$) is treated fully consistently
by respecting detailed balance within the same algoritm.
With this we can adress
the important question of thermalization and early
pressure build up on a partonic level for heavy ion collisions at RHIC.
First (and preliminary) results show that indeed the gluon multiplication
via Bremsstrahlung (and absorption) is of utmost importance,
where thermalization is achievd on a timescale of about 1 fm/c.

\section{Dissipative quantum dynamics for $\phi^4$-theory}

The starting point of our recent numerical investigation 
are the causal Dyson-Schwinger equations
for certain real-time one-particle Green's functions,
which are also called Kadanoff-Baym equations \cite{Ju03}:
\begin{eqnarray}
\label{kabaeqcs}
- \left[
\partial_{\mu}^{x} \partial_{x}^{\mu} \!+ m^2
\right] \, G^{>/<}(x,y)
& = &
\Sigma^{\delta}(x) \; G^{>/<}(x,y) \\[0.5cm]
+ 
\int_{t_0}^{x_0} \!\!\!\!\! dz_0 \int \!\!d^{d}\!z &&
\left[\,\Sigma^{>}(x,z) - \Sigma^{<}(x,z) \,\right] \: G^{>/<}(z,y)
\nonumber
\\[0.5cm]
-
\int_{t_0}^{y_0} \!\!\!\!\! dz_0 \int \!\!d^{d}\!z &&
\Sigma^{>/<}(x,z) \: \left[\,G^{>}(z,y) - G^{<}(z,y) \,\right] \!
\nonumber
\end{eqnarray}

Besides the Hartree term, to next order in the coupling constant 
the non-local sunset self-energy
enters as
\begin{equation}
\label{sunset_cs}
\Sigma^{>/<}(x,y) 
\; = \; - \frac{\lambda^2}{6} \;
\left[ \, G^{>/<}(x,y) \, \right]^3 \, .
\end{equation}\

\begin{figure}[htb]
\begin{center}
\includegraphics[width=9cm]{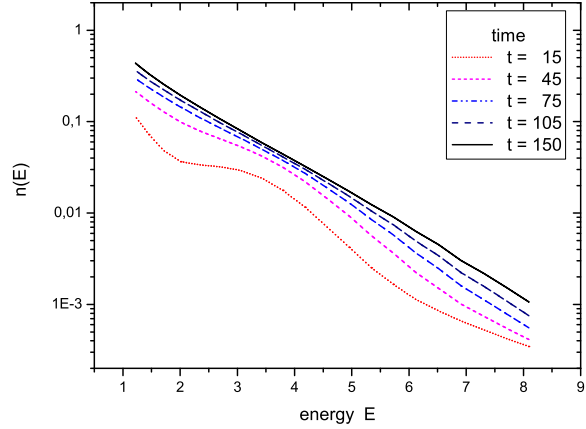}
\end{center}
\caption{\label{slope}
Occupation number spectra along the transverse momentum
axis $(p_x=0, p_y)$ plotted versus (transverse) energy.
The time is taken in units of the inverse mass m, the energy is
given in units of m. }
\end{figure}

With this the Kadanoff-Baym equations are in closed form and can thus
be (numerically) integrated. In the following we summarize 
some major findings of such an analysis for a 2+1 dimensional
and spatially homogenous system being prepared by some non-equilibrium
initial configuration of excited modes in momentum space
at $t=0$
via
\begin{equation}
\label{ew}
2 \omega_{\bf p} \: i\,G^{<}_{\phi \phi}({\bf p},t,t) 
\; = \; 
2 \, n( {\bf p},t) \: + \: 
1 .
\end{equation}
A typical evolution for this Green's function is depicted in 
fig.~\ref{fig:3d1}.
Starting from an initially far non-isotropic shape it develops towards
a rotational symmetric distribution in momentum space. 
Initially there is no occupation along the $p_y$-momentum axis.
As a function of time this area of momentum space is getting
filled which takes about 100 time units. In fig.~\ref{slope}
the occupation number spectra along the transverse momentum axis
is depicted. A quasi exponetial slope is experienced on the very onset
of the ongoing evolution, where the system is still far from kinetic
equilibrium, as most energy is still stored in 
excitations along the longitudinal axis. Hence, a `thermal' spectra
along the transverse direction
as given by the present example does not necessarily imply
that the (complex) system is close to kinetic and thermal equilibrium.
(This could be also true for relativistic heavy ion collisions -
see eg \cite{S03} - where the transverse spectra typically look
`thermal'.)  

\begin{figure}[htb]
\begin{center}
\includegraphics[width=9cm]{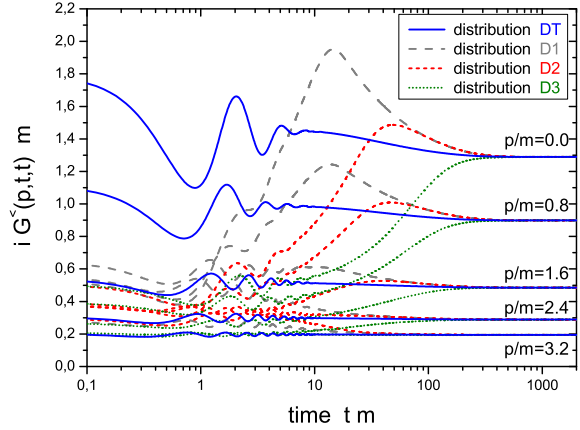}
\end{center}
\caption{\label{fig:equi01}
Time evolution of selected momentum modes of the
equal-time Green function $|\,{\bf p}\,|/m =$ 0.0, 0.8, 1.6, 2.4,
3.2 (from top to bottom) for four different initial
configurations D1, D2, D3 and DT (characterized by the different
line types) with the same energy density.
All momentum modes
assume the same respective equilibrium value for long times
($t \cdot m > 500$) independent of the initial state.}
\end{figure}

On the other hand, we observe that starting from very 
different initial conditions 
for systems containing the {\em same} energy that the single momentum modes  
converge to the same respective numbers for large times as
characteristic for a system in equilibrium. The different momentum modes in
Fig. \ref{fig:equi01} typically show a three-phase structure. For
small times ($t \cdot m < 10$) one finds damped oscillations that
can be identified with a typical switching on effect at $t=0$,
where correlations build up and dephasing of the various modes does occur.
For `intermediate' time scales ($10 < t \cdot m < 100$) one  observes
a strong change of all momentum modes in the direction of the
final stationary state. We address this phase to `kinetic'
equilibration and point out, that the momentum modes can
temporarily even exceed their respective equilibrium value
(see the evolution of the distribution D1). We remark that this behaviour 
of `overshooting' -- as in the
particular case of D1 -- is {\em not} observed in a simulation with a
kinetic Boltzmann equation.
Hence this highly nonlinear effect must be attributed to quantal 
off-shell and memory effects not included in the standard
Boltzmann limit. The third phase, i.e. the `late' time  evolution 
is characterized by a smooth approach of the single momentum modes
to their respective equilibrium values.
As we will now explain,
this phase is adequately characterized by chemical
processes changing the `number' of (quasi-)particles or excitations. 
(It turns out that this rate for chemical equilibration is only slightly
smaller than that of kinetic equilibration for the coupling parameter
chosen \cite{Ju03}.)

\begin{figure}[htb]
\begin{center}
\includegraphics[width=0.6\textwidth]{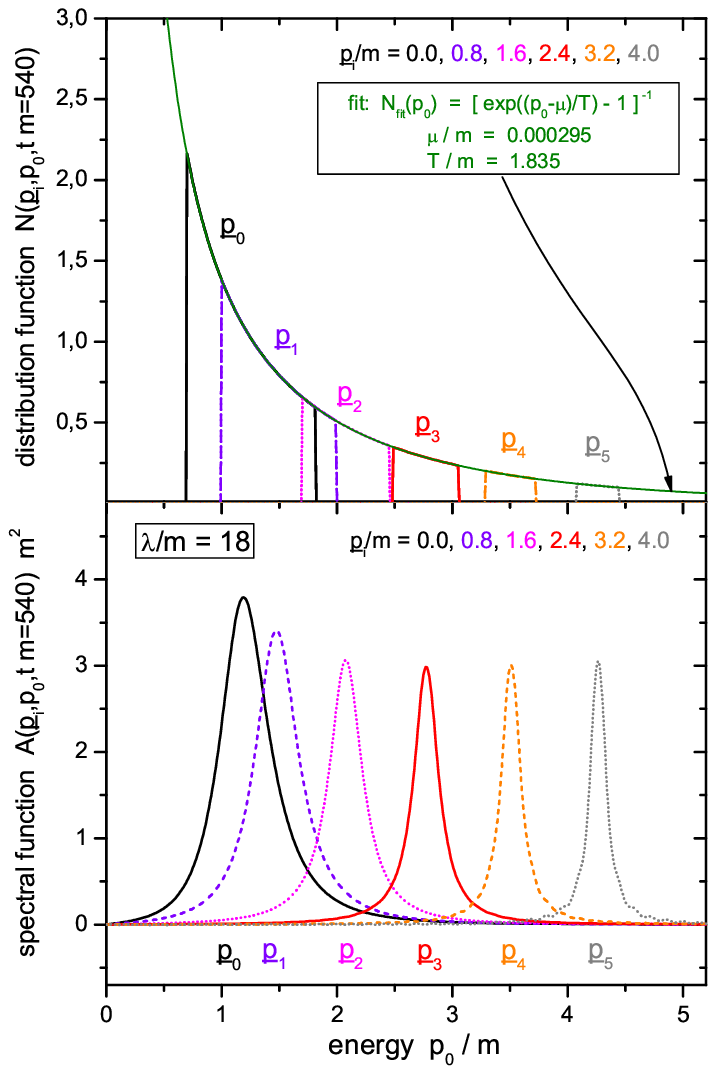}
\end{center}
\caption{\label{fig:kmsna}
Spectral function $A$ for various momentum modes 
$|\,{\bf p}\,|/m =$ 0.0, 0.8, 1.6, 2.4, 3.2, 4.0 
as a function of energy for late times $\bar{t} \cdot m = 540$ 
(lower part). 
Corresponding distribution function $N$ at the same time for the 
same momentum modes (upper part). 
All momentum modes 
with respective off-shell equilibration
can be fitted with a single Bose function of 
temperature $T_{eq} / m = 1.835$ and a chemical potential close 
to zero.}
\end{figure}

In principle one can now study the time evolution of
the spectral function as a function of time.
This is an important and lively topic, but we refer
the interested reader to the paper \cite{Ju03}.
In Fig. \ref{fig:kmsna} (lower part) the spectral function
$A({\bf p},p_0,\bar{t})$ for the initial distribution D2 at very late times
for various momentum modes
is depicted 
as a function of the energy $p_0$.
The system has nearly equilibrated and the spectral function are rather broad,
reflecting a strongly interacting system via the choice
of a rather large coupling constant.

At perfect thermal equilibrium
the famous KMS condition of the Green functions hold dictating that
the equilibrium form of the complete and off-shell distribution function
at temperature $T$ is simply given via
\begin{equation}
\label{distequi}
N_{eq}({\bf p},p_0) \; = \;
N_{eq}(p_0) \; = \;
\frac{1}{\exp(p_0/T) - 1} \; = \;
N_{bose}(p_0/T) \, ,
\end{equation}
which is the well-known Bose distribution.
The distribution function $N(p_0)$ as extracted from the calculation 
is displayed in Fig. \ref{fig:kmsna} (upper part) for the same
momentum modes as a function of the energy $p_0$.
We find that $N(p_0)$ for all individual momentum modes
can be fitted by a single Bose function with temperature
$T/m = 1.835$.
Thus the distribution function emerging from the Kadanoff-Baym
time evolution for $t \rightarrow \infty$ approaches 
indeed a Bose function 
in the energy $p_0$ that is {\em independent} of the momentum as demanded by 
the equilibrium form. The KMS conditions are approached as a dynamical
fixed point in the complex non-equilibrium evolution.

We note that the chemical potential $\mu$
-- used as a second fit parameter -- is indeed close to zero for
these late times as expected for the correct equilibrium state of the
neutral $\phi^4$-theory which is characterized by a vanishing chemical
potential $\mu$ in equilibrium.
This is a consequence 
of the exact treatment.
A kinetic Boltzmann equation in general would lead
to a  stationary state 
for $t \rightarrow \infty$ with a finite chemical potential.

\begin{figure}[htb]
\begin{center}
\includegraphics[width=0.7\textwidth]{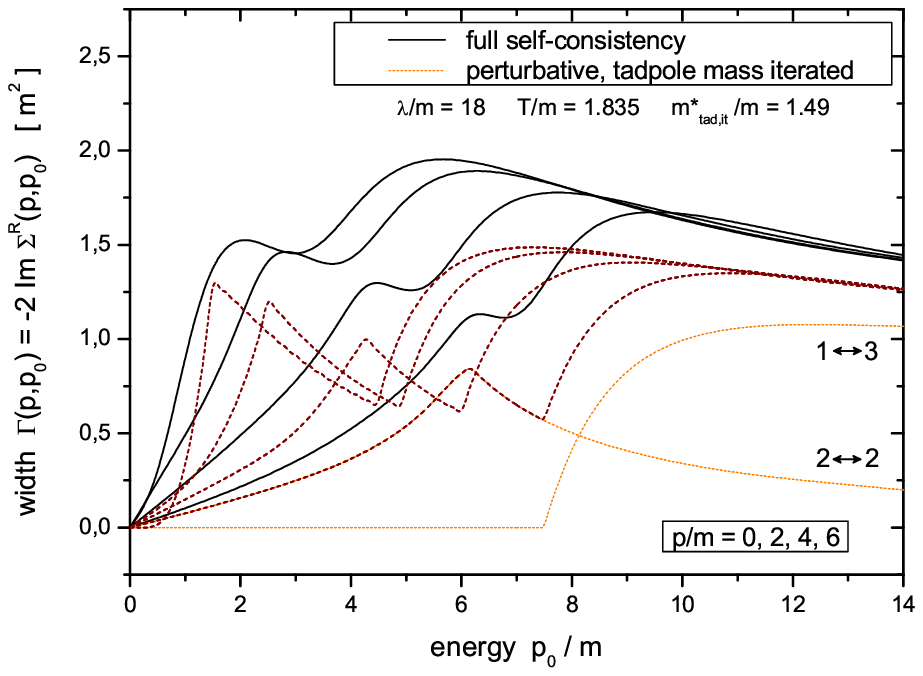}
\end{center}
\caption{\label{fig:selfconsw}
Self-consistent width (solid lines) and perturbative width
(dashed lines) as a function of the energy $p_0/m$ for various momentum 
modes $|\,{\bf p}\,| / m =$ 0, 2, 4, 6 for a thermal system.
For the highest momentum mode $|\,{\bf p}\,| / m = 6$ of the perturbative
calculation, the collision contribution $(2\!\leftrightarrow\!2)$ 
and the decay contribution $(1\!\leftrightarrow\!3)$ to the width 
are explicitly displayed.}
\end{figure}

In Fig. \ref{fig:selfconsw} the width $\Gamma$ is displayed 
as a function
of the energy $p_0$.
The width 
shows two maxima for all momentum modes.
The first is characterized 
by an increase towards a maximum around the 
on-shell energy and falling off beyond.
This behaviour stems from the $2 \leftrightarrow 2$ processes in the
self-energies.
Particles can be scattered by other particles -- present in the 
system at finite temperature -- such that they achieve the shown 
(collisional) damping width.
On the other hand, particles with sufficient energy can decay into 
three other particles.
Above (approximately)
the threshold of $p_{0,th}({\bf p}) = \sqrt{{\bf p}^2+(3\,m^{*})^2}$
these $1 \leftrightarrow 3$ processes lead to an increase of the
width (as marked for the highest momentum mode by the second thin line).
The inclusion of this part of the particle width in the
spectral function is responsible for the process of chemical
equilibration.
Within the Kadanoff-Baym scheme the off-shell 
$1 \leftrightarrow 3$ transitions lead to a violation 
of `particle number'. The physical interpretation of the dynamics is a
consistent 
treatment of Bremsstrahlungs processes in the medium
via the interplay of production, propagation and 
subsequent decay of highly virtual mode
excitations.

\section{Stochastic disoriented chiral condensates}

In this section we give a brief report on our
findings on the stochastic nature of DCC formation and how to
possibly identify their existence experimentally\cite{Xu00}.
This work resulted from an investigation\cite{PRL}, where we
adressed 
the potential likeliness of an instability leading to a
sufficiently large DCC
event during the evolution of a fireball undergoing
a phase transition within the linear $\sigma $-model.

\begin{figure}[htb]
\begin{center}
\includegraphics[width = 8cm]{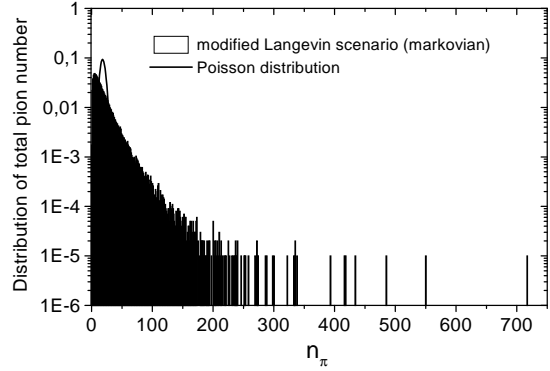}
\end{center}
\caption{\label{DCCfig1}
Statistical distribution $P(n_\pi )$ of the final yield $n_\pi $
in low momentum
pion number of a single DCC for a rapidly expanding
situation (see ref.\protect\cite{Xu00} for details)
compared with a corresponding simple poissonian
distribution.}
\end{figure}

The main idea is that the final fluctuations depend critically on the
initial conditions chosen for the evolving chiral order parameter,
thus deciding to some extent whether the system enters temporarily
the instable region during the `roll-down' period
of the order parameter\cite{PRL,Xu00}.
In fact, a semi-classical and dissipative dynamics of the order parameter
and the pionic fields can be obtained by an effective and complex action,
where the interaction with the thermal pions has been integrated out.
To the end, we have utilized the following stochastic Langevin equations
of motion
for the order parameters $\Phi _a =\frac{1}{V} \int d^3x \,
\phi_a ({\bf x},t)$ in a D-dimensional (`Hubble') expanding volume $V(\tau )$
to describe the evolution of collective pion
and sigma fields\cite{PRL,Xu00}:
\begin{eqnarray}
\ddot{\Phi_0} + \left( \frac{D}{\tau} + \eta \right) \dot{\Phi_0}
+ m_T^2\Phi_0 & = & f_{\pi}m_{\pi}^2 + \xi_0, \nonumber \\
\ddot{\Phi_i} + \left( \frac{D}{\tau} + \eta \right) \dot{\Phi_i}
+ m_T^2\Phi_i & = & \xi_i,
\label{eq1}
\end{eqnarray}
with $\Phi_0=\sigma$ and $\Phi_i=(\pi_1,\pi_2,\pi_3)$
being the chiral meson fields and
$ m_T^2  =  \lambda \left( \Phi_0^2 + \sum_i \Phi_i^2 +
\frac{1}{2} T^2 - f_{\pi}^2 \right) + m_{\pi}^2 $
denotes
the effective transversal (`pionic') masses.
These coupled Langevin equations resemble
in its structure a phenomenological Ginzburg-Landau
description of phase transition.
Aside from a theoretical justification one can  regard the Langevin
equation as a practical tool to study the effect of thermalization
on a subsystem, to sample a large set of possible trajectories
in the evolution, and to address also the question of all thermodynamically
possible initial configurations in a systematic manner.

\begin{figure}[htb]
\begin{center}
\includegraphics[width = 8cm]{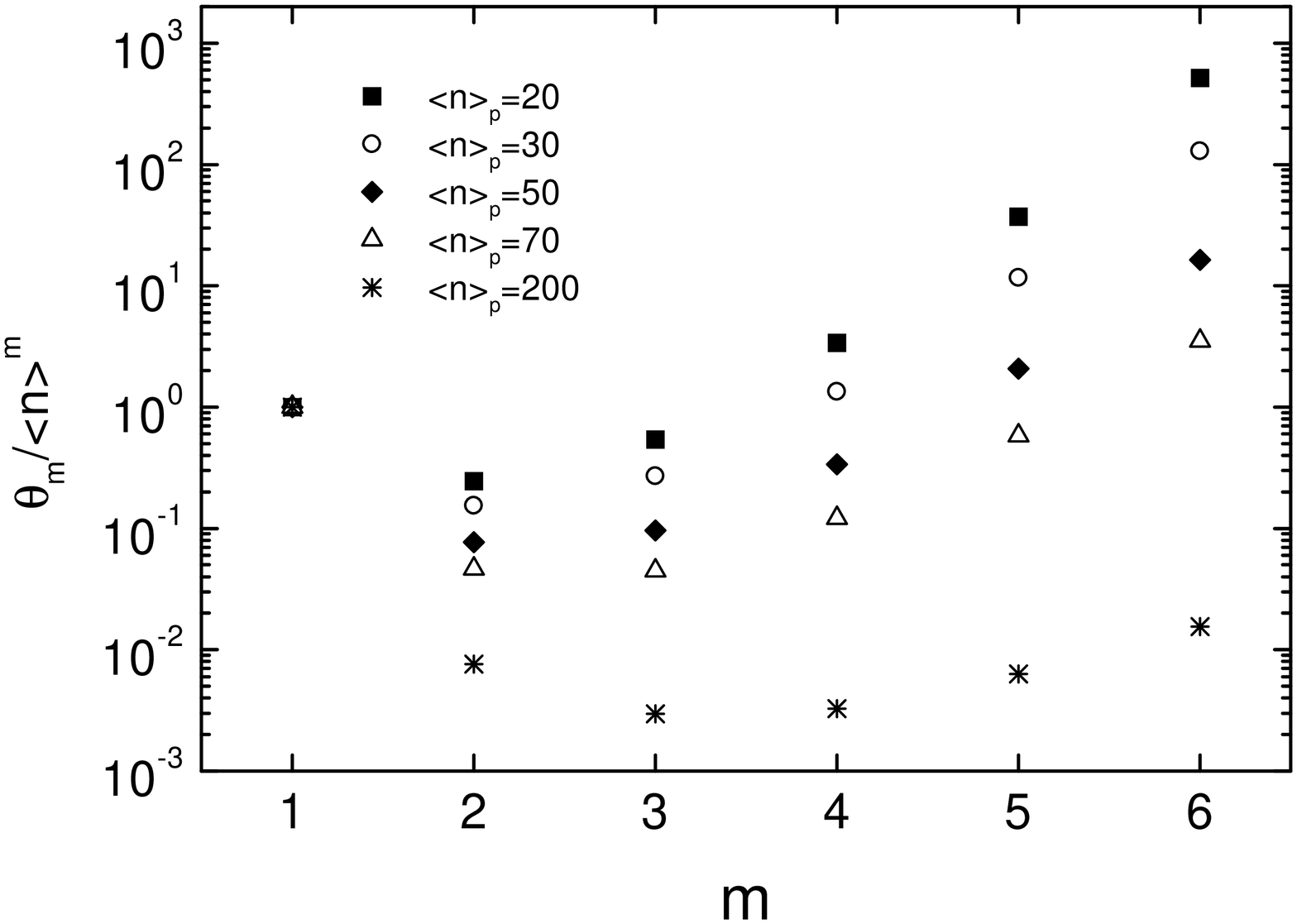}
\end{center}
\caption{\label{DCCfig2}
The reduced factorial cumulants for $m=1$ to 6 for the pion
number distribution of low momentum stemming from
a single emerging DCC (of the previous figure)
and an additional poissonian distributed background pion
source with different mean values $\langle n \rangle_P=20-200$.}
\end{figure}

In fig.~\ref{DCCfig1} we show the statistical distribution in the number
of produced long wavelength pions $N_\pi $ out of the evolving chiral
order fields within the DCC domain $V(\tau )$ for one particular set
of parameters\cite{Xu00}. A rather rapid and ($D=$)3-dimensional expansion
has been employed.
(The results majorly depend on how fast the assumed cooling and expansion
proceeds.)
In general one finds that
only for D=3 and sufficiently fast expansion individual unusual strong
fluctuations of the order of 50 - 200 pions might occur,
although the average number $\langle n_\pi \rangle $ of the emerging
long wavelength pions only posesses a moderate and {\em undetectable} value of 5 -20.

In these interesting cases the
final distribution does {\em not} follow a usual Poissonian distribution
(comp. fig.~\ref{DCCfig1}),
which represents a very important outcome.
(Critical, dynamical) Fluctuations with a large number of
produced pions are still likely
with some small but finite probability!
Unusual events out of sample contain a multiple in the
number of pions compared to the average.
One can interpret those
particular events as semi-classical `pion bursts' similar to
the mystique Centauro candidates.
If DCCs
are being produced, an experimental finding will be a rare event
following a strikingly, nontrivial and nonpoissonian distribution.
A dedicated event-by-event analysis for the experimental
programs (e.g. the STAR TPC at RHIC) is then unalterable.

The further analysis of this unusual 
distribution by means of the cumulant expansion shows that the reduced
higher order factorial cumulants
\\
$\theta_m/<n_{\pi}>^m$ for $m\ge 3$ exhibit
an abnormal, exponentially increasing tendency, as illustrated in
fig.~\ref{DCCfig2}. There an additional incoherent
Poissonian background of (low momentum) pions stemming from other possible
sources has been added.
We advocate that an analysis by
means of the higher order cumulants serves as a powerful signature.
In conclusion, the occurence of a rapid chiral phase transition
(and thus DCCs) can be identified
experimentally by inspecting
higher order facorial cumulants $\theta _m$ ($m\ge 3$)
for taken distributions of low momentum pions.

\section{Elastic and inelastic multi-particle collisions in a parton cascade}

A major goal of the experiments at RHIC and LHC is the creation
and observation of the quark-gluon-plasma.
To describe the dynamics of ultrarelativistic heavy ion collisions (uRHICs),
and to address the crucial question of thermalization and early
pressure build up, we have developed a kinetic parton cascade algorithm
\cite{Xu04}
inspired by perturbative QCD including inelastic (`Bremsstrahlung')
collisions $gg \leftrightarrow ggg $
besides elastic
collisions. It is the aim to get a more detailed
and quantitative understanding of the early dynamical stages
of deconfined matter and to test various initial conditions
for the liberated gluons, than envoking ad hoc
varoius phenomenological, hydrodynamical ansaetze. 
An parton cascade analysis incorporating only 
binary $2 \leftrightarrow 2$
collisions described via direct pQCD scattering processes 
shows that thermalization and early quasi-hydrodynamic behaviour
(for achieving sufficient elliptic flow)
can not be buildt up or maintained,
but only if a much higher cross section is being employed \cite{MG02}.
The possible importance of the inelastic reactions was 
indeed raised
in the so called `bottom up thermalization' picture \cite{B01}.
It is intuitively clear that gluon multiplication should
lead to a faster equilibration.

\begin{figure}[htb]
\begin{center}
\includegraphics[width = 6cm]{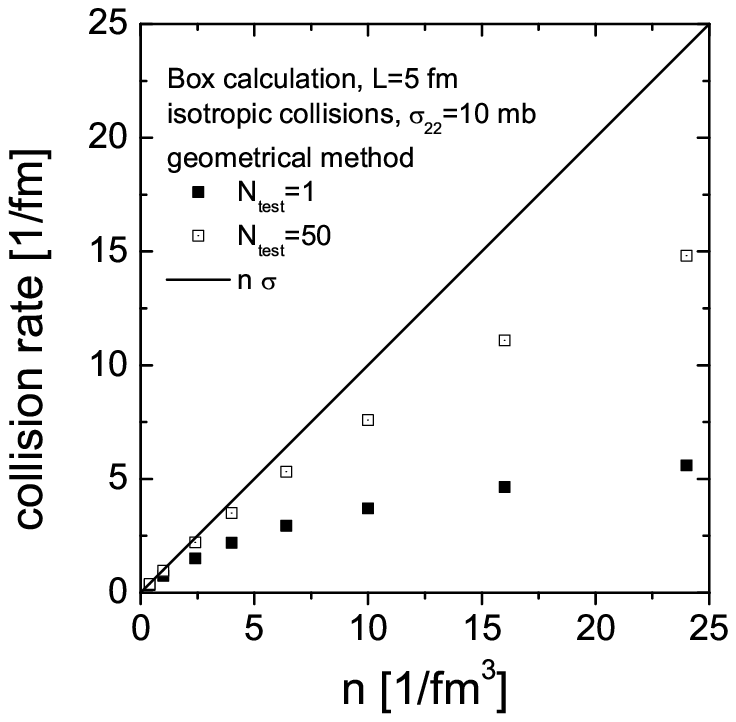}
\vspace*{-0.6cm}
\includegraphics[width = 6cm]{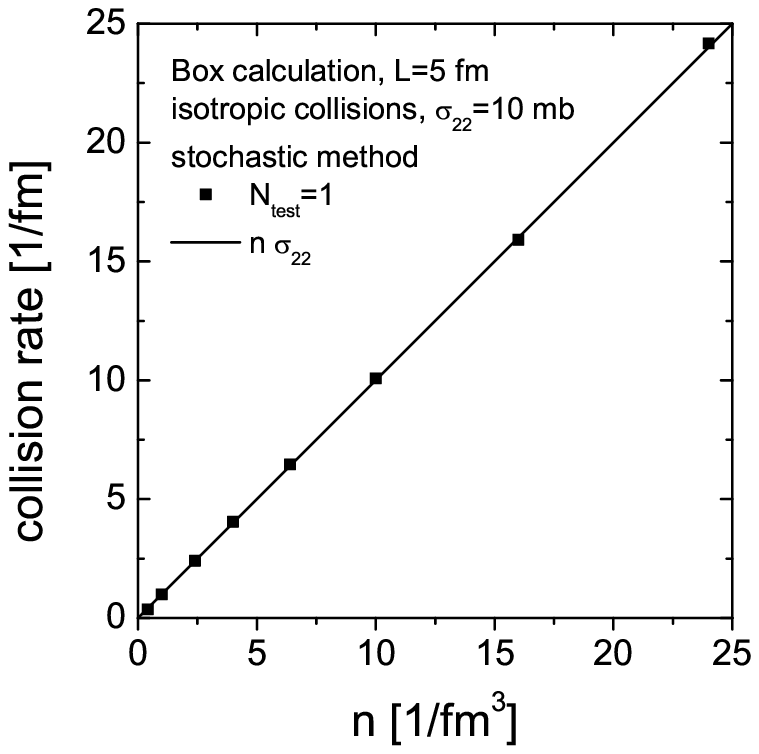}
\end{center}
\caption{\label{Box1}
Collision rates in the simulation versus the ideal rates
$1/(n\sigma ) $. The total cross section is fixed to 10 mb.
In the upper (lower) figure the geometrical (stochastic) method
is employed.}
\end{figure}

In developing the algorithm special emphasis is put on
obeying the principle of detailed balance among the gain and loss
contributions.
The standard incorporation of $2 \leftrightarrow 2$
scattering processes in a transport
description is based on the geometric interpretation of the cross section
\cite{MG00}. For large particle densities,
however, such an implementation leads to
considerable problems to generate a causal collision sequence
among the various partons, leading to a severe reduction of the collision
rate compared to the one dictated by the Boltzmann equation, see
fig.~\ref{Box1}.
In principle, this can be cured by the test particle method.
However, for such a
situation the stochastic method (see eg \cite{La93}) is better suited to
solve the Boltzmann equation directly via transition rates in
sufficiently small spatial cells, see
fig.~\ref{Box1}. Here the actual collision rate being realized in the simulation
fully equals the ideal one of the Boltzmann equation which shall be
simulated. One can go to arbitrary high density.
Moreover, this method can, in principle, be generalized to (any) multi-particle
scattering processes.

\begin{figure}[htb]
\begin{center}
\includegraphics[width = 8cm]{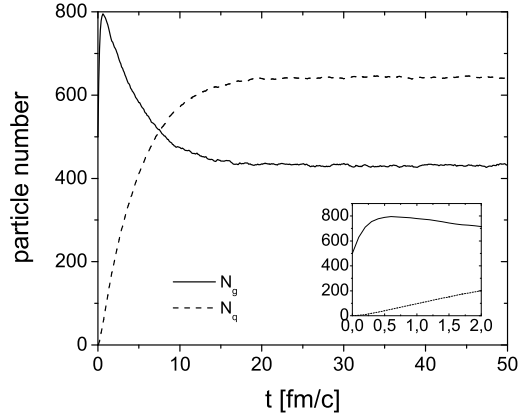}
\end{center}
\caption{\label{Box2}
Time evolution of the gluon and quark numbers for the box calculation.
The gluon number starts at 500.}
\end{figure}

To demonstrate the applicability of the method,
we show in the following the formation of a quark gluon plasma
within a fixed box and
study the way of kinetic and chemical equilibration for different parton
species. The initial conditions of the partons entering the cascade are
given by the multiple minijets (partons with $p_t > 2$ GeV/c) production
\cite{Es89}
from the binary nucleon-nucleon-scattering in a nucleus-nucleus-collision
according to the differential jet cross section:
\begin{equation}
\label{csjet}
\frac{d\sigma_{jet}}{dp_T^2dy_1dy_2} = K \sum_{a,b}
x_1f_a(x_1,p_T^2)x_2f_b(x_2,p_T^2) \frac{d\sigma_{ab}}{d\hat t} \, ,
\end{equation} 
where $p_T$ is the transverse momentum and $y_1$ and $y_2$ are the rapidities
of the produced partons.
The minijets are considered in the central rapidity interval $(-0.5:0.5)$
at RHIC energy of $\sqrt{s}=200$ GeV. The initial partons are distributed
uniformly in a cubic box of size 3 fm,
which approximately corresponds to the central region of heavy ion collions
at some early initial time.

\begin{figure}[htb]
\begin{center}
\includegraphics[width = 8cm]{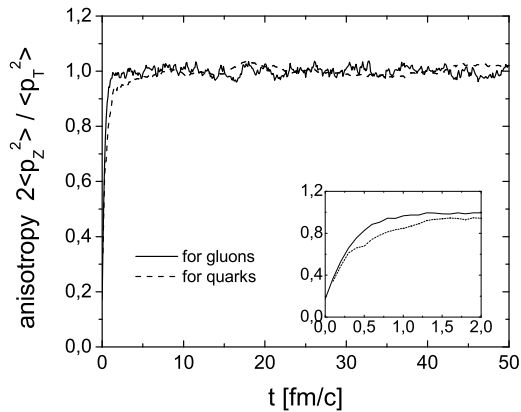}
\end{center}
\caption{\label{Box3}
Time evolution of the momentum anisotropy for the QGP in a box.}
\end{figure}

We consider here gluon and quarks with two flavors as parton species.
Collision processes are the two-body elementary parton-parton scatterings
and three-body process $gg \leftrightarrow ggg $
in leading-order of pQCD,
employing effective Landau-Pomeranchuk-Migdal suppression
and standard screening masses for the infrared sector of the
scattering amplitude.

\begin{figure}[htb]
\begin{center}
\includegraphics[width = 8cm]{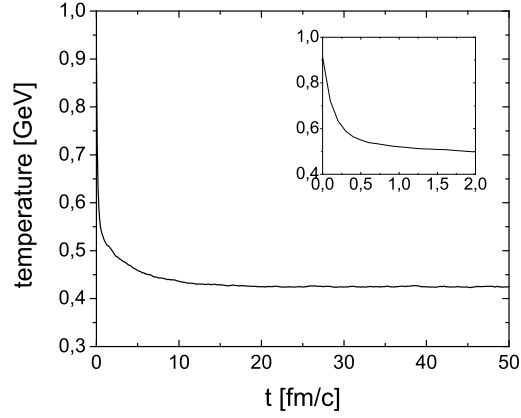}
\end{center}
\caption{\label{Box4}
Time evolution of the effective temperature $T=\frac{E}{3N}$.}
\end{figure}

\begin{figure}[htb]
\begin{center}
\includegraphics[width = 9cm]{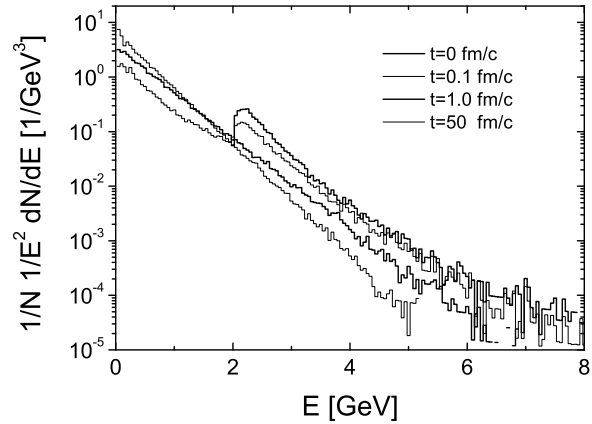}
\end{center}
\caption{\label{Box5}
Occupation versus energy spectra at different times.
within the box calculation.
At $t=0$ only minijets with $E>2$ GeV are populated.
Energy degradation to lower momenta proceeds rapidly by
gluon emission.}
\end{figure}

Fig.~\ref{Box2} shows the time evolution of the gluon and quark number.
Gluon equilibration undergoes two stages: at first the gluon number
increases rapidly and then smoothly evolves to its final equilibrium value
together with the quark number at a much lower rate. 
The early stage of gluon production also leads to an
immediate kinetic equilibration of the momentum distribution
(see fig.~\ref{Box3} and fig.~\ref{Box5})
as well as to a rather abrupt lowering of the
temperature by soft gluon emission until detailed balance
among gain and loss contributions is reached
(see fig.~\ref{Box4}).

The later slower evolution is then governed by
chemical equilibration of the quark degrees of freedom.
The final temperature of fig.~\ref{Box4} is identical
to the late slope parameter
of the energy spectra depicted in fig.~\ref{Box5}.

\begin{figure}[htb]
\begin{center}
\includegraphics[width = 9cm]{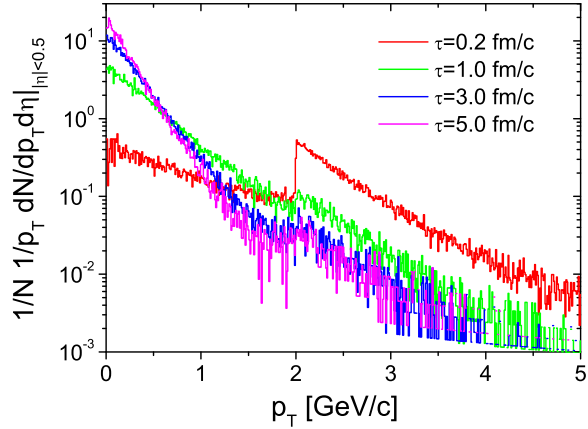}
\end{center}
\caption{\label{minijet}
Transverse momentum spectrum at midrapidity at different times
for a real, central
fully 3-D (including transversal expansion)
ultrarelativistic heavy ion collision. Only the partons residing
in a cylinder of radius $R\leq 5$ fm are taken for the plot.
The initial out-of-equilibrium conditions are given by minijets and
are distributed via the corresponding overlap function in space-time.
The calculation is {\em preliminary}.
From $t=0$ on first only minijets with $E>2$ GeV are populated.
Energy degradation to lower momenta proceeds rapidly by
gluon emission within the first fm/c. Maintenance of
(quasi-)kinetic and chemical equilibrium is given up to 4 fm/c,
where longitudinal and transversal
(quasi-hydrodynamical) work is done resulting in a continous
lowering of the temperature.
On the other hand the initial non-equilibrium high momentum tail
following a standard power-law for mini-jet production is partly
surviving.
}
\end{figure}

With the present cascade  we are aiming to study in detail
real HIC incorporating various initial conditions like minijets
or the color glass condensate. In fig.~\ref{minijet} we depict one
such first (and still {\em preliminary}, as yet not fully tested)
calculation with minijet initial conditions.
Thermalization and chemical equilibration
as propoesed in the bottom up picture \cite{B01} can then thoroghly
be tested within this approach.
Furthermore we will study the impact parameter dependence
of the transverse energy in order to understand elliptic and transverse flow
at RHIC within this new scheme of a kinetic parton cascade including
inelastic interactions by comparing to experimental data.

\begin{acknowledgments}
The work refered to in section 2 has been done together with
W.~Cassing and the collaboration is gratefully acknowledged. 
The various work has been supported by BMBF, DFG and GSI Darmstadt.
\end{acknowledgments}

%

\bibliographystyle{amsunsrt}
\chapbblname{sample} 
\chapbibliography{sample} 


\end{document}